\documentclass{iau} 
\usepackage{graphicx}

\title[IAUS328.~~{Hunting for Stellar Coronal Mass Ejections}] 
{Hunting for Stellar Coronal Mass Ejections}

\author[Korhonen et al.]   
{Heidi Korhonen$^1$, Kriszti{\'a}n Vida$^2$, Martin Leitzinger$^3$, Petra Odert$^{4,3}$, \and Orsolya Eszter Kov{\'a}cs$^{2,5}$}

\affiliation{
  $^1$Dark Cosmology Centre, Niels Bohr Institute, University of Copenhagen, Juliane Maries \\ Vej 30, DK-2100 Copenhagen {\O}, Denmark \\ e-mail: {\tt heidi.korhonen@nbi.ku.dk} \\[\affilskip]
  $^2$Konkoly Observatory, MTA CSFK, Konkoly Thege M. {\'u}t 15-17, 1121, Budapest, Hungary \\[\affilskip]
  $^3$University of Graz, Institute of Physics, Department for Geophysics, Astrophysics and Meteorology, NAWI Graz, Universit{\"a}tsplatz 5, 8010, Graz, Austria \\[\affilskip]
  $^4$ Space Research Institute, Austrian Academy of Sciences, Schmiedlstrasse 6, \\ 8042 Graz, Austria \\[\affilskip]
  $^5$ Department of Astronomy, E{\"o}tv{\"o}s Lor{\'a}nd University, P{\'a}zm{\'a}ny P{\'e}ter s{\'e}t{\'a}ny 1/A, 1117, Budapest, Hungary}

\pubyear{2015}
\volume{xxx}  
\jname{Title of your IAU Symposium}
\editors{A.C. Editor, B.D. Editor \& C.E. Editor, eds.}
\begin{document}

\maketitle

\begin{abstract}
  Coronal mass ejections (CMEs) are explosive events that occur basically daily on the Sun. It is thought that these events play a crucial role in the angular momentum and mass loss of late-type stars, and also shape the environment in which planets form and live. Stellar CMEs can be detected in optical spectra in the Balmer lines, especially in H$\alpha$, as blue-shifted extra emission/absorption. To increase the detection probability one can monitor young open clusters, in which the stars are due to their youth still rapid rotators, and thus magnetically active and likely to exhibit a large number of CMEs. Using ESO facilities and the Nordic Optical Telescope we have obtained time series of multi-object spectroscopic observations of late-type stars in six open clusters with ages ranging from 15 Myrs to 300 Myrs. Additionally, we have studied archival data of numerous active stars. These observations will allow us to obtain information on the occurrence rate of CMEs in late-type stars with different ages and spectral types. Here we report on the preliminary outcome of our studies.
  \keywords{stars: activity, corona, late-type}
\end{abstract}

\firstsection 
\section{Introduction}

Due to the strong radiation coming from hot stars, the cooler stars having later spectral types (F--M) are thought to be more benign hosts for habitable planets. Still, also the cool stars show strong flaring activity and probably numerous coronal mass ejections (CMEs). Magnetic activity of the planet host star, especially the strong magnetic activity exhibited by many rapidly rotating young stars, has profound effects on the planets and their habitability; strong flares and CMEs are thought to even be able to strip a close-by planet of its atmosphere (e.g., \cite[Lammer et al. 2007]{Lammer07}). Additionally, these processes are also relevant for stellar evolution, as stars lose mass and angular momentum via stellar winds and CMEs.

In the Sun CMEs are seen regularly, the average daily occurrence rate during the activity minimum being 0.5 and during the maximum 6, with one solar CME containing on average 10$^{11}$ kg of material (see, e.g., \cite[Gopalswamy et al. 2009]{Gopalswamy09}). On the Sun, flares and CMEs are often closely correlated, their association increasing with flare energy (e.g. \cite[Yashiro et al. 2006]{Yashiro06}). A few studies have therefore aimed to estimate stellar CME rates from flares, and relations between flare and CME parameters known from the Sun (\cite[Aarnio et al. 2012]{Aarnio12}; \cite[Drake et al. 2013]{Drake13}; \cite[Leitzinger et al. 2014]{Leitzinger14}; \cite[Osten \& Wolk 2015]{OstenWolk15}). Many studies have shown that young stars have very energetic Ultra Violet flares (see, e.g., \cite[Osten \& Wolk 2009]{OstenWolk09}), but very few systematic studies of the stellar coronal mass ejections exists. There is a handful of detections of stellar CMEs, but they are very rare (e.g., \cite[Houdebine et al. 1990]{Houdebine90}; \cite[Fuhrmeister \& Schmitt 2004]{FuhrmeisterSchmitt04}; \cite[G{\"u}nther \& Emerson 1997]{GuntherEmerson97}; \cite[Leitzinger et al. 2011]{Leitzinger11}). The observations indicate velocities ranging from about twice the value for very fast solar CMEs (5800km/s seen in AD Leo; \cite[Houdebine et al. 1990]{Houdebine90}) to slightly slower than the values seen in the slowest solar CMEs (84 km/s also seen in AD Leo; \cite[Leitzinger et al. 2011]{Leitzinger11}). The slow velocities observed at times can be explained by the CME being seen in projection.

For their large impact on the environments of planets, and also on the stellar angular momentum evolution, it is crucial to decipher the CME occurrence rate with stellar age and spectral type.

\section{Detecting CMEs using optical spectra}

The Doppler signal of a mass ejection is always seen in projection. It is strongest when plasma is moving towards the observer and weaker for other directions. Doppler shifts and line asymmetries with enhanced blue wings of stellar emission lines can be interpreted as plasma ejected from the star. The material released in CMEs is mainly hydrogen as the CME core is often built by a filament, and they can be detected in the H$\alpha$ line.

The blue enhancement caused by a CME is an absorption feature when seen against the stellar disc, and an emission feature when seen outside the disc. The time that the ejecta can be seen against the stellar disc is quite short for many projections, meaning that emission features would be more common than absorption features. On the other hand, the material would need to be quite dense to be seen in emission. 

The velocities of CMEs span from few hundreds of km/s to some thousands of km/s. The exact value naturally depends on the energetics of the event itself, but also on the projection effects. The maximum velocity is only achieved if the material is ejected directly towards the observer. These high speeds also mean that detecting CMEs does not require very high spectral resolution. 

\section{Results on single stars}

We have gone through the archival data of more than 40 single active stars observed with Narval and ESPaDOnS, and also obtained new observations for a handful of targets. Even with a careful analysis no CMEs were found, except on V374~Peg.

\begin{figure}[t]
\begin{center}
 \includegraphics[width=3in, angle=270]{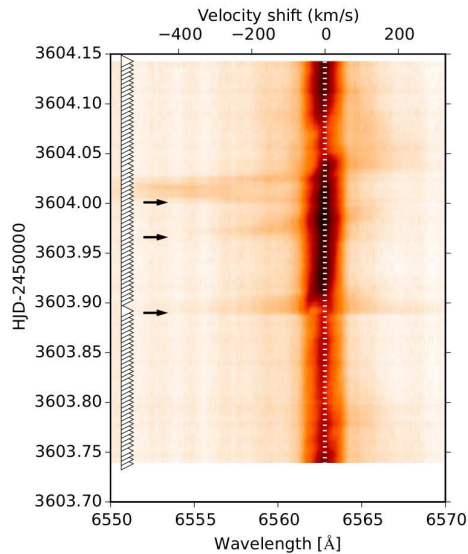} 
 \caption{Dynamic spectrum of H$\alpha$ line of V374~Peg. The arrows show the complex CME event with two failed eruptions and finally one real CME. (from \cite[Vida et al. 2016]{Vida16})}
   \label{V374Peg}
\end{center}
\end{figure}

V374~Peg is a young, active M4 dwarf which has been extensively studied (e.g., \cite[Donati et al. 2006]{Donati06}; \cite[Morin et al. 2008]{Morin08}). We investigated the long-term photometric and spectroscopic variability of this star, and also studied archival ESPaDOnS data of the H$\alpha$ line-profiles (\cite[Vida et al. 2016]{Vida16}). The spectra obtained in 2005 showed a CME event: one real CME that was preceded by two failed events (see Fig.~\ref{V374Peg}). We estimated that the minimum CME mass for the real event was $10^{16}$g. In addition, we predicted the CME rate using the formalism by \cite[Leitzinger et al. (2014)]{Leitzinger14}, and found out that the star should have 15-60 CMEs per day. We detected only one in 10 hours of observations -- clearly less than predicted.

\section{Monitoring of open clusters}

An efficient way for getting a better handle on the CME rate in young stars is to monitor open clusters with different ages using multi-object spectroscopy. In this way many cool stars of known age can be simultaneously observed, increasing the chances of detecting CMEs. 

Our first attempt was done at ESO VLT using VIMOS spectrograph of a young open cluster Blanco~1 (PI Leitzinger). The results from these observations have been published, but unfortunately no CMEs were detected (Leitzinger et al. 2014). Still, we estimated an upper limit of four CMEs per day per star, and that we should have detected at least one CME per star with a mass of $1-15\times 10^{16}$g. After the Blanco~1 campaign we have observed also other young open clusters. IC~2391, NGC~2516, and NGC~3532 were observed using EFOSC2 at ESO's New Technology Telescope (PI Leitzinger). Three further clusters h~Per, IC~348, and NGC~1662 were targeted with the ALFOSC instrument on the Nordic Telescope (PI Korhonen). Unfortunately, many of the NOT observing campaigns were hampered by bad weather, and therefore we decided not to observe the oldest cluster, NGC~1662, which anyway only had few observable cluster members. In total the ALFOSC observations consist of some 2.5 nights worth of good quality data on h~Per and IC~348 (out of 9 allocated nights). More details on the observed clusters is given in Table \ref{clusters}. In this table the observing time is the total time allocated for that cluster.

\begin{table}
  \begin{center}
    \caption{Overview of the clusters observed for this project. Table gives the name of the cluster, age, number of observed targets, instrument used, spectral resolution, typical signal-to-noise ratio (S/N), and the observing time allocated to the cluster. }
    
    \label{clusters}
          {\scriptsize
            \begin{tabular}{|l|l|l|l|l|l|l|l|}\hline 
              {\bf Cluster} & {\bf Age [Myrs]} & {\bf Targets} & {\bf Instrument} & {\bf Resolution} & {\bf Cadence} & S/N & {\bf Obs time} \\ \hline
              Blanco 1 & 65 & 28 & VIMOS, ESO & 2500 & 3 min & 40 & 11h \\ \hline
              IC 2391 & 40 & 7 & EFOSC, ESO & 2500 & 10 min & 40 & 1n \\ \hline
              NGC 2516 & 110 & 8 & EFOSC, ESO & 2500 & 6 min & 40 & 1n \\ \hline
              NGC 3532 & 300 & 9 & EFOSC, ESO & 2500 & 6 min & 40 & 1n \\ \hline 
              h Per & 15 & 17 & ALFOSC, NOT & 2500 & 10 min & 30 & 3n \\ \hline
              IC 348 & 45 & 29 & ALFOSC, NOT & 2500 & 10 min & 30 & 3n \\ \hline 
            \end{tabular}
          }
  \end{center}
\end{table}

From the EFOSC observations of IC~2391, NGC~2516, and HGC~3532 we only detected flares. No CME signatures were seen. Also the h~Per observations obtained with ALFOSC only show some flare events, but IC~348 looks more promising. Our targets in this cluster include stars ranging spectral types G--M. Most of the targets with earlier spectral types have H$\alpha$ in absorption, and the later ones in emission.

\begin{figure}[t]
  \begin{center}
    \includegraphics[width=12cm, angle=0]{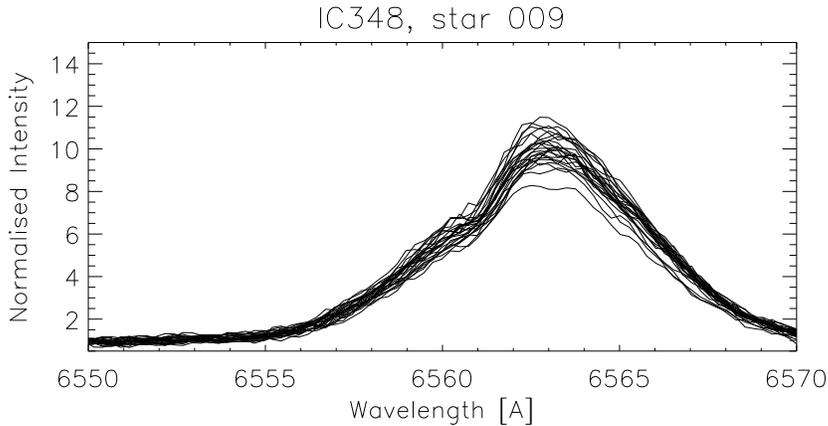} 
    \caption{Examples of H$\alpha$ line profiles of a star in IC~348.}
    \label{IC348_profile}
  \end{center}
\end{figure}

\begin{figure}[t]
  \begin{center}
    \includegraphics[width=6.5cm, angle=0]{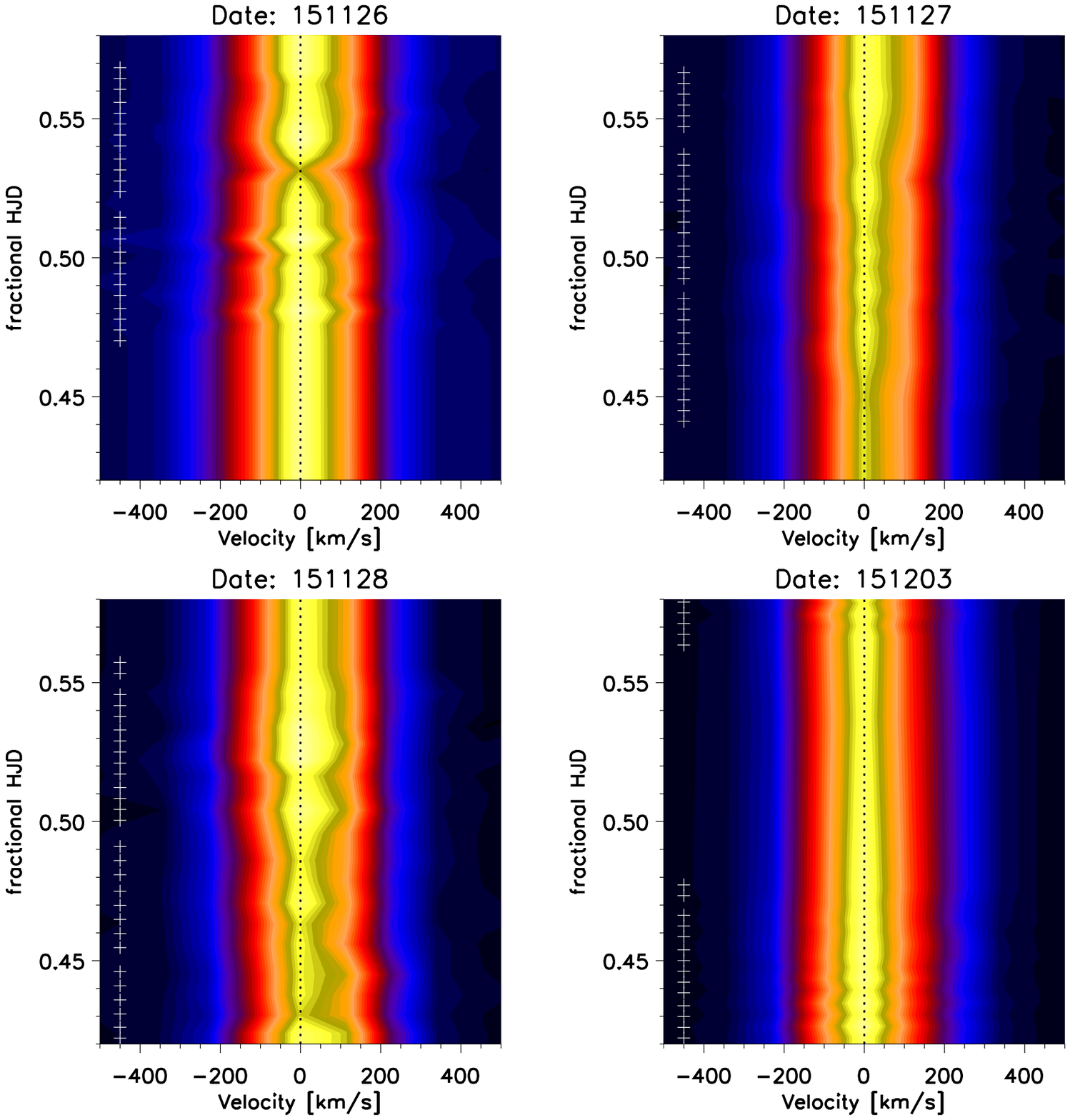} 
    \includegraphics[width=6.5cm, angle=0]{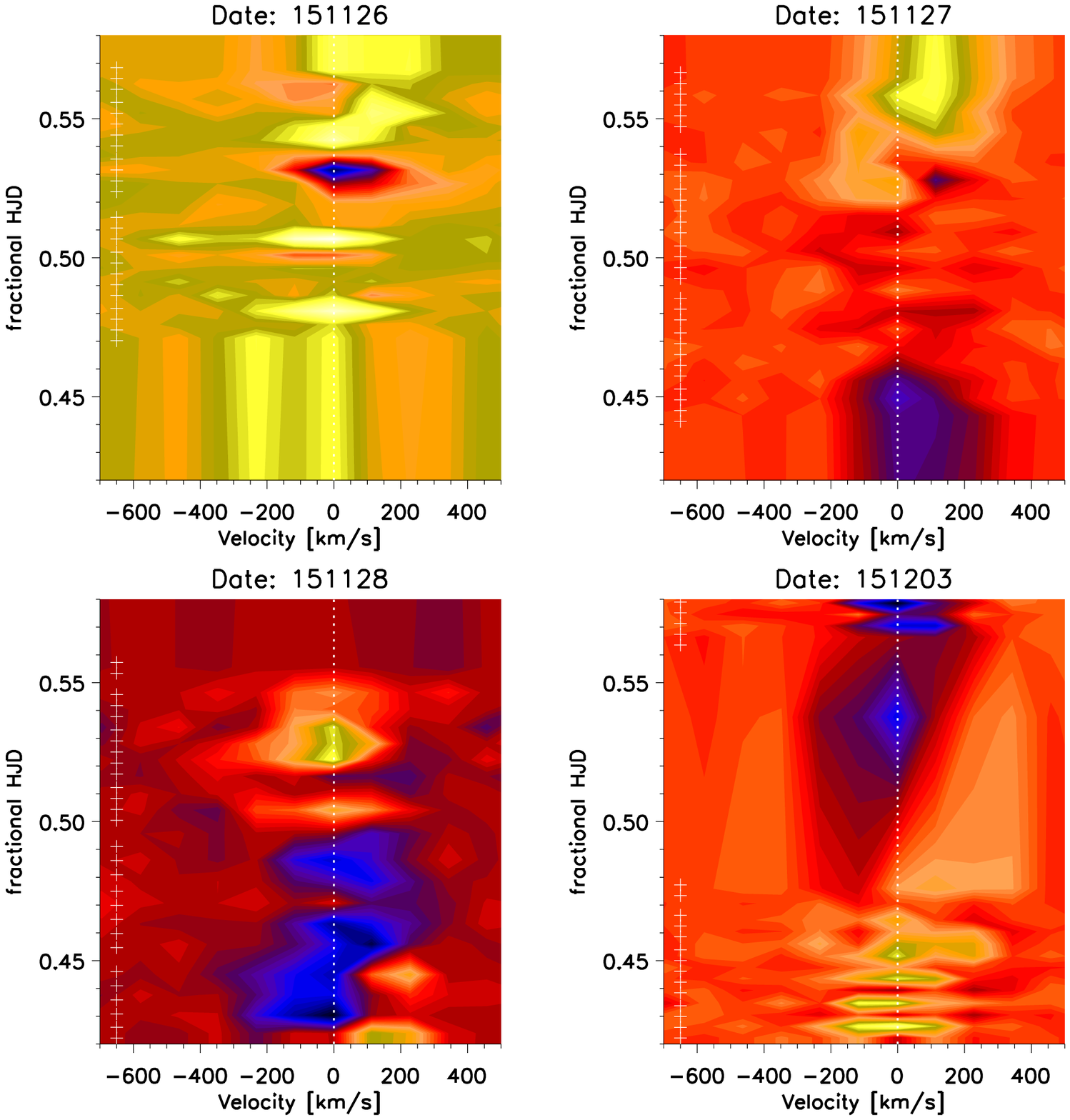} 
    \caption{Dynamic spectrum of H$\alpha$ line of a star in IC~348 (the same star as in Fig.~\ref{IC348_profile}). The observations are from November--December 2015, and the title of the plot gives the exact observing date. The crosses show the times from which the observations are (in the HJD minus HJD of that date at noon). The four leftmost plots show the dynamic spectra from the original profiles, and the four rightmost plots the residual dynamic spectra after the average profile has been subtracted.}
    \label{IC348_dyn}
  \end{center}
\end{figure}

An example of H$\alpha$ line-profiles for star number 9 in our multi-object spectroscopy of IC~348 is given in Fig.~\ref{IC348_profile}. This star is of spectral type late K -- early M. As can be seen, the line-profile shows clear variability. Also, note that the persistent extra emission bump in the blue part of the profile is not from a CME event, but is due to fringing in the CCD at these wavelengths. The dynamic H$\alpha$ spectra for the same target in November--December 2015 are shown in Fig.~\ref{IC348_dyn}. Both the dynamical spectra with original line-profiles (left side) and with the average line-profile subtracted (right side) are shown. Clear variability in the H$\alpha$ region is seen, and several events with blue shifted extra emission are discovered. More thorough analysis of these data are needed for firmly confirming the nature of the events, and calculating their occurrence frequency.   

\section{Conclusions and future prospects}

We are detecting stellar CMEs, but less than expected. The detection is limited both by the intrinsic properties of the star and observing constraints. The intrinsic properties include the CME occurrence rate, which depends on the activity level/age of the star, and the typical CME parameters (velocity, mass, etc.). The observing constraints are on one hand related to the timing of the observations (time coverage and cadence), and on the other to the spectral properties (resolution and S/N).  One has to also keep in mind that a fraction of the CMEs are also lost due to the projection effects.

Typical S/N of our observations is around 30--40, which is not high for studying small features in the spectral line-profiles. With these observations we can detect only the massive CMEs, comparable to the most massive solar events. These events would be in the range of $10^{16}-10^{17}$g. If we increase the S/N then we should also see less massive CMEs. To test this we have been granted UVES bad weather programme on Xi Boo A and B, a 200Myrs old G and K dwarf system. The observations have a resolution of 40000, S/N~600, and time resolution of less than a minute. With these observations we will be able to test the detectability of smaller CMEs.

The so-far detected stellar CMEs have all been on M dwarfs, and the ejecta were found to be in emission. Possibly the CMEs on K and G stars are not dense enough to be detected in emission, and we are not quick enough to detect them in absorption. Another intriguing question is, whether the very active stars really have numerous CMEs. Maybe we are not detecting many CMEs, because there actually are only few of them? It has been hypothesised that the strong magnetic fields on young stars could actually prevent a filament from erupting in analogy to solar failed eruptions (see Drake et al. 2016).

~\\
{\bf Acknowledgments}
H.K.\ acknowledges the support from the {\it Fonden Dr.\ N.P.\ Wieth-Knudsens Observatorium} for the travel grant that made it possible for her to attend the IAU Symposium 328. K.V. was supported by the J{\'a}nos Bolyai Research Scholarship of the Hungarian Academy of Sciences. M.L. and P.O. acknowledge the support from the FWF project P22950-N16. P.O. acknowledges also support from the Austrian Science Fund (FWF): P27256-N27. The authors acknowledge support from the Hungarian Research Grants OTKA K-109276, OTKA K-113117, the Lend{\"u}let-2009 and Lend{\"u}let-2012 Program (LP2012-31) of the Hungarian Academy of Sciences, and the ESA PECS Contract No. 4000110889/14/NL/NDe.

\end{document}